\documentclass[pdflatex,sn-mathphys-num]{sn-jnl}% Math and Physical Sciences Numbered Reference Style 
\usepackage{graphicx}%
\usepackage{multirow}%
\usepackage{amsmath,amssymb,amsfonts}%
\usepackage{amsthm}%
\usepackage{mathrsfs}%
\usepackage[title]{appendix}%
\usepackage{xcolor}%
\usepackage{textcomp}%
\usepackage{manyfoot}%
\usepackage{booktabs}%
\usepackage{algorithm}%
\usepackage{algorithmicx}%
\usepackage{algpseudocode}%
\usepackage{listings}%
\usepackage[utf8]{inputenc}
\usepackage[T1]{fontenc}
\usepackage{bm}

\theoremstyle{thmstyleone}%
\theoremstyle{thmstyletwo}%

\theoremstyle{thmstylethree}%

\begin{document}

\title[Improving ex ante accuracy assessment...]{Improving ex ante accuracy assessment in predicting house price dispersion: evidence from the USA}


\author[1]{\fnm{Adam} \sur{Chwila}}\email{achwila@gmail.com}
\equalcont{These authors contributed equally to this work.}

\author[2]{\fnm{Monika} \sur{Hadaś-Dyduch}}\email{monika.dyduch@uekat.pl}
\equalcont{These authors contributed equally to this work.}

\author[1]{\fnm{Małgorzata} \sur{Krzciuk}}\email{
malgorzata.krzciuk@uekat.pl}
\equalcont{These authors contributed equally to this work.}

\author[1]{\fnm{Tomasz} \sur{Stachurski}}\email{
tomasz.stachurski@ue.katowice.pl}
\equalcont{These authors contributed equally to this work.}

\author[2]{\fnm{Alicja} \sur{Wolny-Dominiak}}\email{
alicja.wolny-dominiak@uekat.pl}
\equalcont{These authors contributed equally to this work.}

\author*[1]{\fnm{Tomasz} \sur{Żądło}}\email{
tomasz.zadlo@uekat.pl}
\equalcont{These authors contributed equally to this work.}

\affil*[1]{\orgdiv{Department of Statistcs, Econometrics, and Mathematics}, \orgname{University of Economics in Katowice}, \orgaddress{\street{50, 1 Maja Street}, \city{Katowice}, \postcode{40-287}, \country{Poland}}}

\affil*[1]{\orgdiv{Department of Statistical and Mathematical Methods in Economics}, \orgname{University of Economics in Katowice}, \orgaddress{\street{50, 1 Maja Street}, \city{Katowice}, \postcode{40-287}, \country{Poland}}}

% \affil[3]{\orgdiv{Department}, \orgname{Organization}, \orgaddress{\street{Street}, \city{City}, \postcode{610101}, \state{State}, \country{Country}}}

\abstract{The study focuses on improving the ex ante prediction accuracy assessment in the case of forecasting various house price dispersion measures in the USA. It addresses a critical gap in real estate market forecasting by proposing a novel method for assessing ex ante prediction accuracy under unanticipated shocks. The proposal is based on a parametric bootstrap approach under a misspecified model, allowing for the simulation of future values and estimation of prediction errors in case of unexpected price changes. The study highlights the limitations of the traditional approach that fails to account for unforeseen market events and provides a more in-depth understanding of how prediction accuracy changes under unexpected scenarios. The proposed methods offers valuable insights for real estate market management by enabling more robust risk assessment and decision-making in the face of unexpected market fluctuations. Real data application is based on longitudinal U.S. data on real estate transactions.}




\maketitle

\section{Introduction}\label{sec:Introduction}

The real estate market behaviour is highly sensitive to long-term fluctuations caused by structural changes. They can result from various political, legal, environmental, technological, and social disturbances currently affecting the market, but also from the dynamic and interconnected nature of these factors. However, as discussed in \cite{perera2017improving}, the complexity of these changes can even eliminate the predictability of the performance of the real estate market in the case of standard forecasting approaches. Therefore, it is crucial to incorporate into price dispersion forecasting process potential shocks that have not been observed or measured in the past.
In \cite{leeper2013fiscal} market shocks resulting from fiscal policy are classified into two categories based on their future occurrence: anticipated and unanticipated shocks. However, changes in fiscal policy are among the few economic phenomena that can provide clear signals about how key characteristics of the market will evolve in the future. Because shock influencing the real estate market can result from various reasons and can be very complex,  we adopt the same classification but in the following context. For further analysis, we define anticipated shocks as those assumed in forecasts to occur in a future period. Unanticipated shocks will be defined as those not assumed in forecasts to appear in the future period. It is important, that we do not specify how this information is incorporated into the prediction method. For instance, it can be represented as a binary variable, where a value of one indicates shock periods, as the dependent variable with the appropriate lag, or as an independent variable that is correlated with the impact of the shock. However, an additional classification regarding the estimation of shock effects should be introduced to clarify the approach which will be proposed in this paper. We distinguish three types of shocks: shocks with impact estimable based on sample data, estimable based on external sources and inestimable. Joint discussion of these two classifications leads to the following findings. 
Firstly, if a shock is estimable using sample data, it can become the anticipated shock, and we can also estimate the ex ante prediction accuracy.  Secondly, when a shock is estimable only through external sources, it can remain classified as anticipated, but estimating ex ante prediction accuracy becomes not trivial. The complexity of this problem arises because the information about the estimated impact of the shock, which is necessary for forecasting, is insufficient to assess the ex ante prediction accuracy. Finally, if a shock is inestimable, it can be classified as unanticipated, which means that its impact cannot be incorporated into forecasts. This situation may occur even if information on real estate prices during a shock is available, but when the estimation of the shock effect based on the current data is not possible. For example, if data on real estate prices lacks geolocation information, it would be impossible to measure how flooding affected the prices of properties exposed to the event. Another case is law changes that affect a specific subpopulation of real estate (e.g., properties with certain legal or technical attributes), which may be present in the data but remain unidentifiable due to the lack of a variable that specifically captures these properties. However, as will be proposed in this paper, the prediction accuracy in such situations can still be estimated not only based on a standard model without shock impacts but also based on scenarios covering this type of shocks. Summing up, in the paper, we focus on the shocks which are inestimable or estimable only based on external sources and their influence on the ex ante prediction accuracy.

% przegląd literatury o uwzględnianiu szoków przez pryzmat wprowadzonych klasyfikacji szoków - część 1
Let us discuss the approaches presented in the literature for incorporating market shocks into the analysis based on the classifications introduced. 
Typically, information about shocks in a market is incorporated into models estimated based on sample data as additional independent variable or variables. An interesting example of a shock estimable based on sample data is presented in \cite{cohen2021storm}, where a real estate market shock caused by a hurricane is modelled using four unit-level variables: a \textit{post-storm} variable (0-1), which equals 1 if a property is sold after the storm; \textit{exposure}, which measures the distance from the storm surge; \textit{surprise}, defined as the difference between the actual storm surge and what was expected based on floodplain maps; and \textit{exposure - floodplain distance}. Similar considerations are presented by \cite{holtermans2024quantifying} and \cite{petkov2022weather}. This approach is usually used to estimate and test the effects of shocks, but it also allows the prediction of future values of prices, taking into account the estimated impact of these shocks in the forecasts for the future period (leading to anticipated shocks estimable based on sample data according to our classification). This problem is addressed in \cite{maboudian2020effect}, which analyses the impact of four macroeconomic shocks on the GDP growth rate using historical data and computes forecasts for future periods. In \cite{lin2021minimizing}, a slightly different approach is proposed. Effects of shocks estimated based on external sources are used to compute forecasts of share prices, but the accuracy assessment is limited to the ex post analysis. Summing up, this approach can only be applied when the current sample or external data allows for the quantification of the shock's impact on the market. In this paper, we focus on situations where this condition does not have to be met. 

% przegląd literatury o uwzględnianiu szoków przez pryzmat wprowadzonych klasyfikacji szoków - część 2
%TZ: rozważam połączenie tego akapitu z powyższym akapit.
Moreover, in the literature shock scenarios are incorporated into simulation models, usually to analyse the changes in prices over time in such situations.  An example of such a study, which generates prices to mimic changes during the periods of Great Moderation followed by the Global Financial Crisis, is presented in \cite{meen2022long}. 
%In \cite{miao2014discount}, discount-rate shocks as a link between commercial real estate and the business cycle are of primary interest. 
In \cite{miao2014discount} not only the generation of real estate data under shock scenarios but also the analysis of predictive power (as it is called by the Authors) is discussed. However, this analysis is limited to the comparison of two measures for models estimated based on real data and model-simulated data: (a) $R^2$ values and (b) the values of the estimators of one of the model parameters. Although this approach is slightly similar to our own, as the Authors also consider inestimable shocks (they assume a specific realisation of the shock in the future), it does not cover the estimation of ex ante prediction accuracy, which is of primary interest in this paper.

Our aim is to improve the prediction process by providing additional information on ex ante prediction accuracy in the context of unanticipated market shocks. 

\section{The improved method of prediction assessment} \label{sec:the_improved}
We assume that the available forecasts do not take into consideration any information about market shocks, or take into account the anticipated shocks. Then, the analysis is conducted under three scenarios: first, when there are no shocks (or when anticipated shocks occur); second, when unanticipated shocks happen in the market; and third, when unanticipated shocks occur with a defined probability. It is crucial to emphasize that we are not attempting to predict unanticipated shocks, nor are we simulating prices to compare models with and without unanticipated shocks, as previously discussed. Our approach covers the following steps.
\begin{enumerate}
\item Prediction of dispersion measures without shocks (or with anticipated shocks)
\item Improved ex ante assessment of prediction accuracy of the forecast covering:
    \begin{enumerate}
    \item estimation of ex ante prediction accuracy under the assumed model (without shocks or with anticipated shocks), 
    \item estimation of ex ante prediction accuracy under misspecified model: 
    \begin{itemize}
    \item under scenarios with unanticipated shocks occurring with a specific probability,
    \item under scenarios with unanticipated shocks occurring with probability 1.
    \end{itemize}
    \end{enumerate}
\item Comparing the results obtained under unanticipated shock scenarios (see point 2(b)) with reference values (see point 2(a)).
\end{enumerate}
%Our approach to accuracy prediction estimation in case of shocks on the market differs in both theoretical aspects and practical applicability from the two approaches presented above. 
We consider a situation when information about a shock is not available or when it is not possible to identify the influence of a shock based on the existing sample data. Hence, we are not able to compute forecasts taking the shock into account. However, we can incorporate assumed shock scenarios when assessing ex ante prediction accuracy. This can be accomplished using parametric bootstrap under what is referred to as a misspecified model (e.g., \cite{ghosh2012asymptotic}). This model enables the generation of sample data mimicking the real observations for previous periods and, at the same time, does not have to mimic past price changes in the data generation process for the future period. The proposed approach allows the simulation of values of forecasts $\hat\theta$  based on real sample data (without identified shocks) and the generation of future values of predicted quantities $\theta$ under (assumed to be unexpected) shock scenarios at the same time. Therefore, the prediction errors $\hat\theta - \theta$ cover the unknown future scenarios. It implies that the proposed method allows for estimating ex ante prediction accuracy under any future scenario, including unanticipated shocks and the occurrence of unanticipated shocks with assumed probabilities, which extends and improves the assessment of forecasting methods. This addresses a significant gap in existing forecasting approaches, where predictions often fail to consider the potential impact of unforeseen market events on their accuracy. Finally, the proposed approach can help to understand how the ex-ante prediction accuracy of the chosen prediction method is changing if the scenario not taken into account in the prediction process will appear in the future.

Although we model real estate prices based on longitudinal data, the aim is to predict their dispersion measures, which are of significant interest in real estate analyses (e.g., \cite{kwong2000price}, \cite{leung2006housing}, and \cite{kanno2022land}). We consider various metrics to effectively capture price variability during both periods when shocks are observed and those when they are not. Moreover, our analysis extends beyond the entire USA, addressing different subpopulations within the U.S. real estate market. This can provide an additional informative perspective for assessing regional social inequality. In \cite{stachurski2024predicting} pp. 988-989, quantile-based measures were used to measure the dispersion of real estate prices, including interquartile range ($IQR$), quartile dispersion ratio ($QDR$), decile dispersion ratio ($DDR$), and quartile coefficient of dispersion ($QCD$). However, especially when market shocks appear, researchers should aim to capture the full extent of price dispersion. Therefore, we have expanded our analysis by incorporating several classic variability measures. These include the standard deviation ($SD$), the average absolute deviation ($AAD$), and the coefficient of variation ($CV$). Including these measures allows for more comprehensive variability analyses, capturing heavy tails and outliers, which frequently occur during periods of market shocks.

%The real estate market is sensitive to fluctuations resulting from structural market changes which may impact it even for years or decades. Therefore, possibility of incorporating shocks that have not been observed or their influence has not been measured in the past into price dispersion forecasting is crucial for the market analyses. It is due to the significant climatic, political, and economic changes characterizing the current economy and the dynamic and interconnected nature of these factors. 

% DLACZEGO MODELE MIESZANE

The most popular classes of models used for analysing real estate data cover machine learning techniques such as deep learning methods \cite{jiang2021modeling}, gradient boosting decision trees \cite{han2022financial}, neural networks \cite{fong2013prediction}, and support vector machines \cite{yang2024data}. However, a limitation of these models is that their usage implies that the prediction accuracy is assessed through the ex post analysis, such as k-fold cross-validation. As a result, their application in the considered scenarios involving unobserved or non-quantifiable shocks in the real estate market is significantly constrained. This limitation is avoided for parametric regression models. In the case of modelling real estate data, the multiple linear regression method is widely used (e.g., \cite{kim2013predicting} and \cite{liu2022research}). Other models, including evolutionary polynomial regression \cite{tajani2019multivariate}, structural vector autoregression \cite{guinea2023news}, and geographically weighted regression in \cite{chrostek2013spatial}, are also considered. We propose using linear mixed models, also utilized in \cite{cichulska2018analysis} and \cite{arribas2016mass} (but not under market shocks), because of two reasons. This class of models can capture complex spatio-temporal dependence structures often observed in longitudinal data and, at the same time, allows for ex ante assessment of prediction accuracy under the proposed approach.

The ability to assess how inestimable shocks affect prediction accuracy can lead to the improvement of the selection of the forecasting method. Hence, it enables researchers and practitioners to make the choice more robust on various market scenarios. Moreover, it can be treated as a proposal of a more complex and flexible framework for understanding and predicting real estate market behaviour in the case of various uncertainties. The layout of this paper is as follows. In Section \ref{sec:USAmarket_dispersion} the considered dispersion measures are introduced and their values for the considered longitudinal data are presented. The proposed method is discussed in Section \ref{sec:improved} and its application in Section \ref{sec:application}. Finally, the conclusion of this study is presented in Section \ref{subsec:Conclusion}.

\section{USA market -- dispersion analysis} \label{sec:USAmarket_dispersion}
Although there are many papers analysing the US real estate market, including the assessment of the variability in the past periods, they usually overlook the prediction of dispersion measures for the future period or the estimation of their ex ante prediction accuracy. For example, in \cite{lisi2013real} the price dispersion understood as the phenomenon
of selling two houses with very similar attributes and locations at the
same time but at very different prices, is considered but it is not predicted. Similarly, in \cite{plazzi2008cross} the cross-sectional dispersions of returns and growth in rents for commercial real estate based on longitudinal data from US metropolitan areas are explored, but again, no forecasts of dispersion measures are presented.

Even if the problem of prediction is considered, it is usually limited to the prediction of real estate prices for the future period, not their dispersion measures. Moreover, even in the case of prediction of property prices, the ex ante assessment of prediction accuracy, crucial for effective decision-making, is not taken into account as in \cite{gupta2022machine} and \cite{walacik2024property}. An exception can be found in results presented in \cite{stachurski2024predicting}. However, the Authors do not consider the variability of property prices (but the variability of the sum of the prices in various regions), their analyses do not take into account shocks on the real estate market, do not encompass all the variability measures considered in this paper, and  do not study the US real estate market.

In this section, various dispersion measures are introduced  and their behaviour in past periods is discussed. In further sections, the problem of their prediction will be discussed as well as the proposed methods for estimating their ex ante prediction accuracy, including scenarios involving unanticipated market shocks. 

\subsection{Dispersion measures}\label{subsec:measures}
This section presents various measures of dispersion and inequality applied to evaluate the variability in the real estate market. Forecasting volatility requires an appropriate statistical approach. On the one hand, the use of dispersion measures serves as a fundamental tool in risk analysis, enabling the prediction of the impacts of various market shocks. On the other hand, it is thought that social inequality, especially income inequality, and price volatility in the real estate market are related. This conclusion is based on patterns observed across multiple OECD countries, where recent decades have seen a systematic increase in property prices alongside rising social inequalities. Numerous studies demonstrate that this is not a coincidental dependence but rather a causal one \cite{Goda2021}. The purpose of this study is, therefore, to assess the degree of price dispersion in the real estate market, which the authors assume to be correlated with the prevalence of social inequalities. Similar concepts have been explored in research using quantile-based measures of dispersion \cite{stachurski2024predicting}. This study extends this approach by incorporating classical measures and selected robust metrics of dispersion. 

The measures considered in the paper can be categorized into four groups: classical absolute measures of dispersion, positional absolute measures of dispersion, relative measures of dispersion, and measures of inequality.
The group of classical absolute measures of dispersion includes the \textbf{Standard Deviation (SD)} and the \textbf{Average Absolute Deviation (AAD)}, defined by the following formulas \eqref{SD} and \eqref{AAD}:
    \begin{equation} \label{SD}
        SD = \sqrt{\frac{1}{n-1} \sum_{i=1}^n (y_i - \bar{y})^2},
    \end{equation}
    \begin{equation} \label{AAD}
        AAD = \frac{1}{n} \sum_{i=1}^n |y_i - \bar{y}|.
    \end{equation}    
These measures provide insight into the degree of variation in property values and form a foundation for assessing market heterogeneity. However, it should be noted that these measures could be insufficient to capture the extent of dispersion, especially for highly skewed distributions \cite{leung2006housing}. 

Another approach is based on robust methods, which employ positional measures.  This group of methods, referred to as positional measures of dispersion, comprises the \textbf{Interquartile Range (IQR)} and the \textbf{Median Absolute Deviation (MAD)} given by \eqref{IQR} and \eqref{MAD}:
\begin{equation} \label{IQR}
	IQR = q_{0.75} - q_{0.25},
\end{equation}
where $Q_1$ and $Q_3$ are the first and third quartiles, respectively.
    \begin{equation} \label{MAD}
        MAD = Me (|y_i - \text{Me}(y)|),
    \end{equation}
Use of such methods can be profitable, especially in the case of asymmetric distribution. These measures are also used in risk analyses \cite{Sehgal2023}.

Another method of assessing dispersion is based on relative measures. This is particularly advantageous when comparing distinct subgroups, such as regions or segments of a market. This third group of methods called relative measures of dispersion consists of the \textbf{Coefficient of Variation (CV)} and the \textbf{Quartile Coefficient of Dispersion (QCD)}  given by the following formulas:
    \begin{equation}\label{CV}
       CV = \frac{SD}{\bar{y}}\cdot100\%.
    \end{equation}
    \begin{equation} \label{QCD}
        QCD = \frac{Q_3 - Q_1}{Q_3 + Q_1}\cdot100\%,
    \end{equation}
The coefficient of variation is commonly used in analysis of the price dispersion \cite{Chiang2019} or comparison income dispersion between states in USA \cite{Nissan2001}. 
The Quartile Coefficient of Dispersion is considered a suitable substitute for the CV. Unlike the CV, it always takes values between 0 and 100\%.

The final category encompasses measures of inequality based on quantile ratios, such as the \textbf{Decile Dispersion Ratio} and \textbf{Quartile Dispersion Ratio}, as outlined below.
\begin{equation} \label{QDR}
	QDR=\frac{Q_{3}}{Q_{1}},
\end{equation}
where $Q_{3}$, $Q_{1}$ are quartiles, respectively, the third and the first.
\begin{equation} \label{DDR}
	DDR=\frac{D_{9}}{D_{1}},
\end{equation}
where $D_{9}$, $D_{1}$ are deciles, respectively, the ninth and the first.
In inequality analysis, a commonly used measure is the Gini coefficient, which, however, is not considered in this study. As an alternative, we suggest the use of dispersion ratios. Dispersion ratios are particularly effective at capturing changes in the tails of the distribution, unlike the Gini coefficient. The Gini coefficient is highly sensitive to changes in the center of the distribution \cite{Jedrzejczak2021}.

\subsection{Dispersion -- USA real estate market}

We analyze the dispersion of house prices using a longitudinal dataset consisting of 25 561 sale transactions from the US real estate market between 2015 and 2023. It contains information regarding the real estate market and its macroeconomic environment, and it will be discussed in Subsection \ref{subsec:model}.

\begin{figure}[ht] 
\centering
\includegraphics[scale = 0.55]{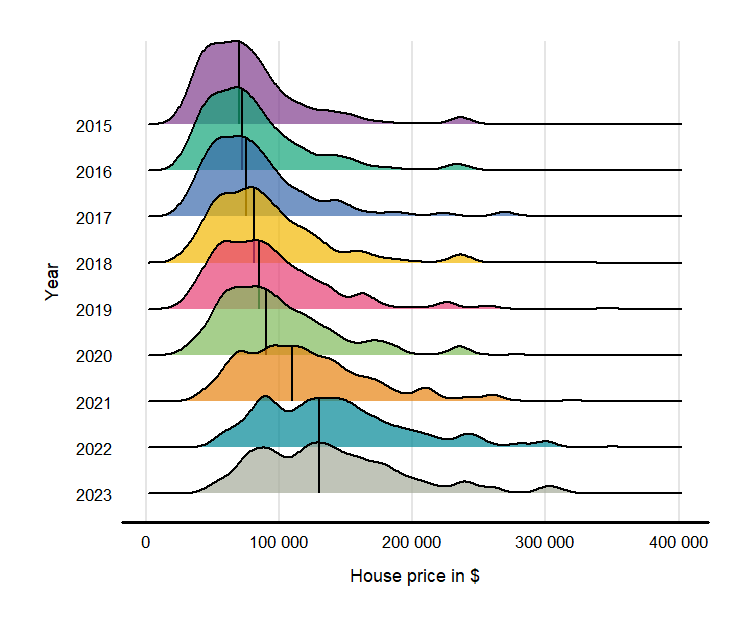}
\caption{Distributions of house prices in the USA in years 2015--2023}
\label{prices_9years}
\end{figure}

The changes of the distributions of prices over the years in the entire US real estate market are illustrated in Figure \ref{prices_9years}. The vertical lines on the empirical density functions represent the medians, which have increased over time, indicating a significant shift toward higher prices. In each period a positive skewness is observed, as well as the multimodality of the distributions. This situation requires a detailed analysis of the dispersion.

%These transactions span across 7 years, reflecting a comprehensive temporal coverage. In addition, the data set includes data from five unique regions, providing a diverse geographic perspective for analysis.

% \begin{figure}[ht]\label{dist}
%     \centering
% \includegraphics[scale = 0.3]{dist.png}
% \caption{The left panel : distribution of House Prices in years. The right panel: distribution of House Prices in years on logarithmic scale. }
% \end{figure}

% Figure \ref{dist} illustrates the distribution of house prices on two different scales: linear and logarithmic. The left panel shows the distribution of house prices in US dollars. This distribution is strongly right-skewed, indicating that the majority of house prices are concentrated in the lower range, while a smaller number of houses exhibit very high prices, resulting in a long tail of the distribution. The median house price is $89,000$, which reflects the value that divides the data set into two equal parts, with half the prices below and half above this value. The skewness of the distribution is evident, as the median is significantly lower than the likely mean, which is influenced by the presence of high-value outliers. Applying a logarithmic transformation greatly reduces the skewness of the data, making the distribution more symmetric and resembling a normal distribution. Together, the two graphs demonstrate the importance of logarithmic transformation in handling highly skewed data.

%Analysing the dispersion measures in 2023, the values show moderate variability with an SD of 0.404, while the AAD (0.324) and MAD (0.268) suggest lower dispersion, indicating reduced sensitivity to extreme values. The IQR of 0.582 reflects a moderate spread in the central 50\% of the data. A low CV of 0.034 implies minimal relative variability. The QCD (0.025) suggests very low dispersion relative to the median, whereas the QDR (1.051) and DDR (1.093) indicate relatively high dispersion in the middle and extreme data ranges, potentially suggesting skewness or outliers.

We consider values of dispersion measures not only in the whole population, but also within subpopulations defined by two variables:
\begin{itemize}
\item REGION - four census regions: Northeast, Midwest, South, West, and a fifth code that represents the national level for three or more section homes,
\item SECTIONS  - the size of the home: single, double, three or more section.
\end{itemize}
To consider subpopulations in the following years is essential for a nuanced understanding of the dispersion on the housing market, which can vary significantly across different geographical areas, housing types and time. As presented in \cite{hwang__2006}, the US real estate market is not homogeneous and it can be treated rather as a collection of diverse regional markets, each affected by local economic, demographic, and regulatory factors. Furthermore, studies indicate that this regional specificity changes over time, leading, for example, to region-specific bubble shocks (\cite{miles_2015} and \cite{freese_2015}) and varying across regions herding behaviors that increase price dispersions \cite{ngene2017time}. Furthermore, analyses focusing on the whole US market may not account for interactions between regions, such as the transmission of housing market shocks between regions studied in \cite{antonakakis_2019}.  

The considered subpopulations, defined by all combinations of categories of variables REGION and SECTIONS available in the dataset, are as follows:  
\begin{itemize}
    \item subpopulation 1:  one section houses in Northeast region,
    \item subpopulation 2:  one section houses in Midwest region,
    \item subpopulation 3:  one section houses in South region,
    \item subpopulation 4:  one section houses in West region,
    \item subpopulation 5:  two section houses in Northeast region,
    \item subpopulation 6:  two section houses in Midwest region,
    \item subpopulation 7:  two section houses in South region,
    \item subpopulation 8:  two section houses in West region,
    \item subpopulation 9:  three (or more) section houses (on the whole US market).
\end{itemize}

\begin{figure}[ht!] 
    \centering
\includegraphics[scale = 0.75]{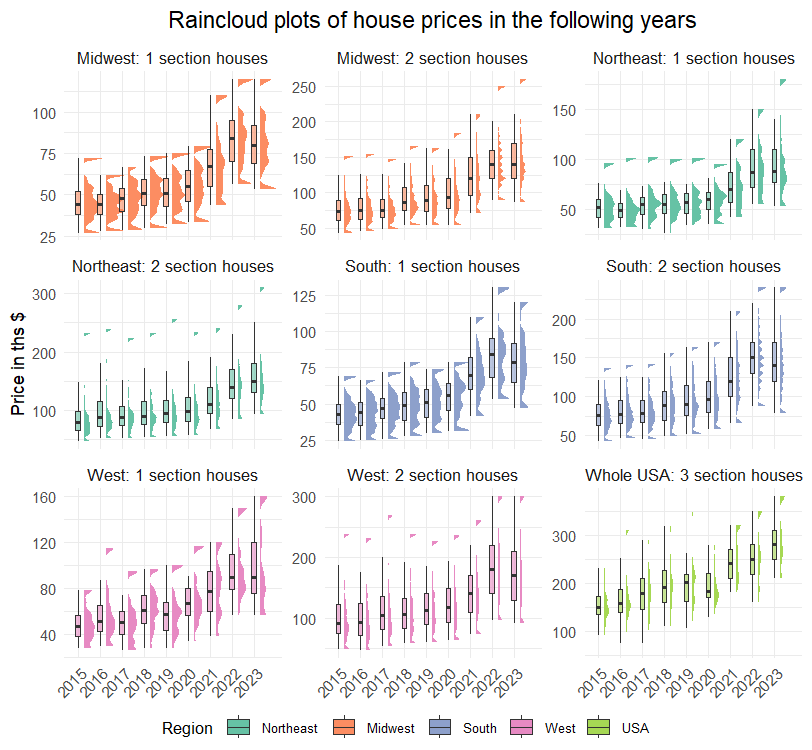}
\caption{Distributions of house prices in subpopulations in years 2015--2023}
\label{prices_subpop}
\end{figure}

The changes in price distributions over the years across the considered subpopulations are illustrated in Figure \ref{prices_subpop}. 

This figure shows the changes between the subpopulations as well as the changes in time within each subpopulation, including a rapid increase of prices after 2020 in several subpopulations. Various ranges of the prices across the considered subpopulations confirm the necessity of using relative dispersion measures. Moreover, the presence of a large number of outliers suggests the need to use both classical and quantile-based measures depending on whether the variability analysis aims to account for these outliers or maintain robustness against them.

\begin{figure}[ht!]
    \centering
\includegraphics[scale = 0.85]{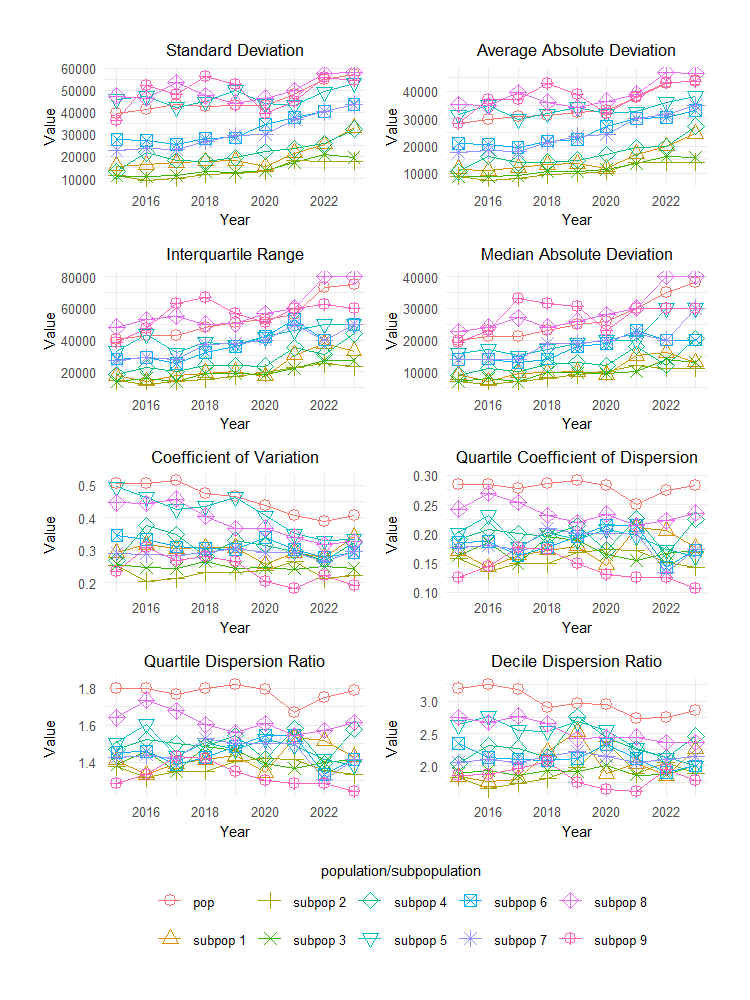}
\caption{Values of dispersion measures in the whole population and subpopulations in years 2015--2023}
\label{dispersion_measures}
\end{figure}

Based on the considered historical data values of the dispersion measures introduced in Subsection \ref{subsec:measures} are computed for the entire USA and the subpopulations of interest over the studied years. Their values are presented in Figure \ref{dispersion_measures}. 

The variability within the population and subpopulations evolves over time. These changes differ both across various measures and among different subpopulations. For example, most subpopulations exhibit relatively stable results over time for relative measures, while we observe increasing trends in standard deviation and average absolute deviation. Similarly, different subpopulations display varying levels of variability. For instance, subpopulation 9 consistently shows the highest variability, whereas subpopulations 1 and 3 maintain lower levels of dispersion. Additionally, in some subpopulations, such as subpopulation 4, increased fluctuations in certain variability measures, like dispersion ratios, are noted after specific periods (e.g., after 2020). As a result, the dynamic nature and complexity of these changes lead to different trends across subpopulations, diverse behavior of the considered measures, and emerging disparities at certain times, which greatly complicates the forecasting process.

\section{Theoretical background of the improved method of prediction assessment}
\label{sec:improved}

In Section \ref{sec:the_improved}, we present a three-point procedure for the improved method of prediction assessment. In this section, we introduce the theoretical background of each step in three subsections: \ref{subsec:procedure_step1}, \ref{subsec:procedure_step2}, and \ref{subsec:procedure_step3}.

\subsection{Prediction of dispersion measures in the absence of unanticipated shocks} \label{subsec:procedure_step1}

Let $\mathbf{Y}_T$ be a random vector representing the logarithms of real estate prices in a future period $T$. Our aim is not to predict $\mathbf{Y}_T$ but rather to predict their function given by $\theta(G^{-1}(\mathbf{Y}_T))$, where $G^{-1}(.)$ is a back-transformation function (here: $exp(.)$ due to the logarithmic transformation of prices) and $\theta(.)$ in our case is one of the dispersion measures (\ref{SD})-(\ref{DDR}). To predict $\theta(G^{-1}(\mathbf{Y}_T))$ a model should be assumed for both future and past values of logarithms of real estate prices. The assumed model can account for information on a shock, allowing for the estimation of the shock impact on the prices and including this information as the anticipated shock in the forecast.

% \textcolor{red}{We consider the problem of prediction of any given function of the population vector  of the response variable: ..under the LMM. It covers linear combinations of (such as one future realization of the response variable or population and subpopulation means and totals) but also other population and subpopulation characteristics such quantiles and variability measures.}

This section discusses the prediction of dispersion measures of real estate prices in the future period based on real data from the U.S. market. Firstly, we will introduce the considered dataset and the proposed longitudinal mixed model. Secondly, we will present the predictor of dispersion measures under the model, where real estate prices (not dispersion measures) are the dependent variable. The issue of estimation of ex ante prediction accuracy of the considered predictor under unanticipated shocks will be considered in the next section.

% \textcolor{red}{
% Krótki opis jak robimy predykcję Z CYTATEM DO SIR \cite{SIR2024}. 
% Napisać, że tu tylko predykcja bez estymacji dokładności predykcji (to w następnej sekcji). TU predyktor nie uwzględnia uanticipated, inestmable shocks (które są przedmiotem naszego zaintersowania), co jednak nie wyklucza, że model może uwzględniać zmienne objaśniające pozwalające na uwzględnienie innych szoków - tych anticipated.}

\subsubsection{Prediction model} \label{subsec:model}

%The first problem is - independent variables ....\\

%czy my to wyjaśniamy ostatecznie?
% To explain the changes in logarithms of prices (LOG.PRICE), we initially consider over 350 auxiliary variables obtained as discussed in Subsection \ref{subsec:data}. The choice of the final combination of the independent variables is made through the following steps:

%bezpieczniej:
To investigate the factors driving changes in the logarithms of prices (LOG.PRICE), we start by examining a comprehensive set of over 350 auxiliary variables. This data set has been prepared by integration various data sources, including the American Community Survey, the Manufactured Housing Survey, data from the Bureau of Labor Statistics, and information from the Board of Governors of the Federal Reserve System. The selection procedure, the final combination of independent variables and their description are presented in Appendix \ref{App.A}.

%%%%%%%%%%%% Poniższy tekst jest na początku tej sekcji w ogólniejszej formie 
%%%%%%%%%%%% (PROSZĘ NIE KASOWAĆ, MOŻE SIĘ PRZYDA)
%The information on HOUSEHOLDS, LOG.OWNER, and INCOME is sourced from the American Community Survey available on the U.S. Census Bureau website. Data on PPI is obtained from the U.S. Bureau of Labor Statistics website, while information on FED comes from the Board of Governors of the Federal Reserve System website. All other variables are derived from the Manufactured Housing Survey, also provided by the U.S. Census Bureau website.

Let us consider the final mixed model assumed for the sample vector of random variable LOG.PRICE denoted by $\textbf{Y}_S$. For the $i$th element of $\textbf{Y}_S$ it can be written as follows:
\begin{equation} \label{themodel}
    Y_{idt} = \mathbf{x}_{idt}^T {\bm{\beta}} + {v}_{id} + {e}_{idt},
\end{equation}
where $i = 1, 2, \dots, n$, $d = 1, 2, \dots, D$, $t = 1, 2, \dots, M$, $n = 25 561$ sale transactions, $D = 5$ regions, $M = 9$ years, $Y_{idt}$ is the logarithm of the price of the $i$th property, $\mathbf{x}_{idt}$ is a vector of the values of the independent variables for the $i$th property (described in Appendix \ref{App.A}), ${\bm{\beta}}$ is the vector of unknown fixed effects,  ${v}_{id}$ is the region-specific random effect, and ${e}_{idt}$ is the random components, $v_{id}$ and $e_{ikt}$ are mutually independent. Taking into account region-specific random components allows to include spatial and temporal correlation of prices within regions. The Restricted Maximum Likelihood method is used to estimate model parameters, which provides consistent estimators even under the lack of normality. 

It is important to note that modelling house prices based on the considered dataset is challenging due to the following reasons. The main reason is the lack of strong correlations between house prices and various independent variables. As an illustration, Figure \ref{scatterplots} in Appendix \ref{App.B} shows scatter plots of house prices and one of these variables, specifically house areas, across all considered subpopulations and years. Moreover, the dataset lacks comprehensive auxiliary information, such as detailed geolocation data for each property, which means that it is not possible to quantify the impact of the shocks considered later in the article on property prices. %However, these challenges are common in various market analyses, especially when dealing with market shocks.

\subsubsection{Prediction method} \label{subsection:Prediction_method}

The typical problem considered in the literature is to predict future property prices. In such case, we aim to predict an element or elements of the population random vector $\mathbf{Y}_T$ of logarithms of real estate prices in the future period $T$. The predictor of $\mathbf{Y}_T$ vector will be denoted by $\hat{\mathbf{Y}}_T$. Its $i$th element which belongs to the $d$th region under (\ref{themodel}) is given by 
\begin{equation} \label{yhat}
\hat{Y}_{idT}=\mathbf{x}_{idT}^T {\hat{\bm{\beta}}} + \hat{v}_{id},    
\end{equation}
%AWD%%A nie powinno być \widehat E[YidT]??
where ${\hat{\bm{\beta}}}$ is the sample estimator of ${\bm{\beta}}$, $\hat{v}_{id}$ is the Empirical Best Linear Unbiased Predictor of $v_{id}$ (see \cite{rao_small_2015} p. 108), and $\mathbf{x}_{idT}^T$ is the assumed vector of independent variables for the $i$th property in the future period. However, the problem discussed in this paper is not the prediction of prices of specific properties but the prediction of dispersion measures in the whole population (as a function of $\mathbf{Y}_T$) or in a subpopulation (as a function of a subvector of $\mathbf{Y}_T$).  

Hence, the considered prediction problem can be defined as a general problem of prediction of $\theta(G^{-1}(\mathbf{Y}_T))$, where $G^{-1}(.)$ is the back-transformation function (here: $exp(.)$ because of the log transformation of prices), and in our case $\theta(.)$ will be replaced by measures (\ref{SD}) -- (\ref{DDR}). In order to predict $\theta(G^{-1}(\mathbf{Y}_T))$, we can employ a versatile method known as the PLUG-IN predictor, which can be used with any parametric or nonparametric model. This method is discussed in detail by \cite{boubeta2016empirical}, \cite{hobza_empirical_2016}, and \cite{chwila_properties_2022}. As demonstrated, for instance, by \citet[][p.~20]{chwila_properties_2022}, the PLUG-IN predictor may offer greater or comparable accuracy when compared to the Empirical Best Predictor. The Empirical Best Predictor is an estimated version of the predictor that minimises the prediction Mean Squared Error (MSE). The PLUG-IN predictor of $\theta(G^{-1}(\mathbf{Y}_T))$ can be expressed as $\hat{\theta} = {\theta}(G^{-1}(\hat{\mathbf{Y}}_T))$. 

However, the usage of this predictor requires the knowledge or assumption of values of all independent variables for each property in the future period denoted in (\ref{yhat}) by $\mathbf{x}_{idT}$. It is worth noting that independent variables HOUSEHOLDS, LOG.OWNER and INCOME are variables lagged by one year, which means that their real values for the future period are known. In the case of monthly values of PPI and FED, it is assumed that their values remain unchanged compared to data from the last available period. For the rest of the independent variables, which are on the property level, we employ a similar strategy as in the case of the Emiprical Best Predictor used in \cite{chwila_properties_2022} and \cite{molina_small_2010}. In the case of unknown values of independent variables for non-sampled population elements, the Authors replicate their values a number of times depending on the sampling weights attached to the sampling units. We use the same approach based on the values from the most recent available period and assume these values will also apply to the forecast period. Therefore, we expect the structure of the real estate market in the upcoming period to be similar to that of the last period. Therefore, the assumed number of future transactions, denoted by $N_T$, is the sum of these weights rounded to the nearest integers.

\subsection{Improved ex-ante prediction accuracy assessment} \label{subsec:procedure_step2}

There are two standard approaches for assessing prediction accuracy: the ex post and the ex ante approach. In cases where shocks are not observed in sample data, the ex post prediction accuracy assessment cannot account for changes in the distribution of the response variable resulting from these events. In the case of the ex ante accuracy estimation without the improvement which will be proposed, the models are also estimated based on sample data where shocks are not observed. Therefore, we proposed an improved method of ex-ante prediction accuracy assessment to cover the influence of unanticipated shocks on prediction accuracy. The improvement is possible by treating the shocks as random variables and by generating the values of the dependent variable, taking their distributions into account. We use the proposed bootstrap under -- so-called -- misspecified model (a model different from the one estimated based on real sample data, e.g., \cite{ghosh2012asymptotic}). As a result, the estimates of the ex ante prediction accuracy measures are obtained, which values take the shock scenarios into account. This topic will be further explored in Subsection \ref{subsec:extended_approach}.

Similar to \cite{stachurski2024predicting}, we recommend assessing prediction accuracy using the following two ex ante prediction accuracy measures. However, they will be used in the proposed extended approach, including the proposed ex ante assessment of prediction accuracy under unanticipated shocks. We can use the well-known prediction RMSE, given by the following formula:
\begin{equation}\label{RMSE}
RMSE(\hat{\theta})=\sqrt{E(\hat{\theta}-\theta)^{2}}.
\end{equation}
It is worth noting that the RMSE is the root of the MSE, the Mean Squared Error. The MSE solely measures only the average of the positively skewed distribution of squared errors. However, we are interested in the comprehensive description of this distribution, not only the mean, especially if unanticipated shocks are to be taken into account. As we are also interested in the relationship between the error magnitude and the probability of its occurrence, we propose employing the QAPE, which is the $p$th quantile of the absolute prediction error $|U|$, as discussed by \cite{zadlo_parametric_2013} and \cite{wolny-dominiak_bootstrap_2022}. It is given by:
\begin{equation}\label{QAPE}
	QAPE_p(\hat{\theta}) = \inf \left\{ {x:P\left( {\left| {{\hat{\theta} - \theta}} \right| \le x} \right) \ge p} \right\}
\end{equation}
This measure informs that at least $p100$\% of observed absolute prediction errors are smaller than or equal to $QAPE_p(\hat{\theta})$, while at least $(1-p)100\%$ of them are greater than or equal to $QAPE_p(\hat{\theta})$. In our opinion, it is a very good measure in the case of shock scenarios influencing the tails of the distribution of prices, allowing us to measure the accuracy by focusing on the influence of shocks.

In practice, it is not feasible to compute (\ref{RMSE}) and (\ref{QAPE}). However, they can be estimated based on any model based on real sample data or generating realization of real estate prices under any shock scenario. 

\subsubsection{Assessment of the ex ante prediction accuracy under the specified model} \label{subsec:traditional_approach}
 
 We use parametric bootstrap instead of the residual bootstrap used for example in \cite{stachurski2024predicting} because it offers greater flexibility for the modifications needed when incorporating shock scenarios. The data generation model will cover two models for property prices. The first model, discussed in this subsection, is designed for a scenario without unanticipated shocks, although it can cover anticipated shocks. The second model, proposed in the following subsection, addresses scenarios involving unanticipated market shocks.
 
 In the first case, the sample realizations of logarithms of prices, denoted by $\textbf{Y}_S$, are generated to receive its bootstrap realizations denoted by $\textbf{Y}^{*}_S$. The $i$the element of $\textbf{Y}^{*}_S$ is generated as follows:
\begin{equation} \label{bootmodel_1S}
    Y_{idt}^{*}= \mathbf{x}_{idt}^T \hat{\bm{\beta}} + {v}^{*}_{id} + {e}^{*}_{idt},
\end{equation}
where $\hat{\bm{\beta}}$ is the estimated, based on sample data, value of $\bm{\beta}$, ${v}^{*}_{id}$ and ${e}^{*}_{idt}$ are generated from the assumed distribution random effects and random components, respectively, $i = 1, 2, \dots, n$, $d = 1, 2, \dots, D$, $t = 1, 2, \dots, M$, $n$ is the number of sale transactions available in the sample longitudinal data, $D = 5$ regions, $M = 9$ years, $\mathbf{x}_{idt}$ is a known vector of the values of the independent variables for the $i$th property presented in Subsecion \ref{subsec:model}.

Future realizations of logarithms of prices in period $T$, denoted by $\textbf{Y}_T$, are generated to obtain its bootstrap realizations $\textbf{Y}^{*}_T$ under the same model. For the $i$th element of  $\textbf{Y}^{*}_T$,  it can be written as follows:
\begin{equation} \label{bootmodel_1T}
    Y_{idT}^{*}= \mathbf{x}_{idT}^T \hat{\bm{\beta}} + {v}^{*}_{id} + {e}^{*}_{idT},
\end{equation}
where $\mathbf{x}_{idT}$ are values of auxiliary variables assumed for future transactions and $i=1,2,...,N_T$, where $N_T$ is the assumed number of future transactions (as described in Subsection \ref{subsection:Prediction_method}). and the rest of the notations are discussed in (\ref{bootmodel_1S}).

Estimators of ex ante prediction accuracy measures $RMSE$ and $QAPE$ under correctly specified are given by equations (\ref{RMSE}) and (\ref{QAPE}), where $\theta=\theta(G^{-1}(\textbf{Y}_T))$ 
and $\hat{\theta}=\hat{\theta}(G^{-1}(\textbf{Y}_S))=\theta(G^{-1}(\hat{\mathbf{Y}}_T))$ are replaced with  
$\theta^{*} = \theta(G^{-1}(\textbf{Y}^{*}_T))$ and $\hat{\theta}^{*} = \hat{\theta}( G^{-1}(\textbf{Y}^{*}_S))=\theta(G^{-1}(\hat{\mathbf{Y}}^{*}_T))$, respectively, where $\hat{\mathbf{Y}}^{*}_T$ is a vector of forecasts computed based on formula (\ref{yhat}) which is based not on real values of logarithms of prices, but on their bootstrapped values $\textbf{Y}^{*}_S$.

Hence, it is assumed that the model generating future prices is the same as the model generating prices for sample data (compare (\ref{bootmodel_1S}) and (\ref{bootmodel_1T})). This situation could cover both no shocks and anticipated shocks, with effects estimated based on sample data presumed to take place in the future period.  

\subsubsection{Extended approach - assessment of the ex ante prediction accuracy under the misspecified model} \label{subsec:extended_approach}

In this subsection it is assumed that sample realizations of the dependent variable follow model (\ref{themodel}), which means that their boostrap realizations can be obtained based on (\ref{bootmodel_1S}). However, future unanticipated shocks are taken into account in bootstrapping for the future period. To cover many possible models with  unanticipated shocks, that differ from the sample model (i.e. various misspecified models), very general bootstrap model generating future realization of independent variables is assumed. Future realizations of $\textbf{Y}_T$ are generated to obtain $\textbf{Y}^{(shock)*}_T$ with the $i$th element given by:
\begin{equation} \label{bootmodel_shock}
Y_{idT}^{(shock)*}= h^{(shock)} \left( 
\begin{bmatrix}
\mathbf{x}_{idT}^T  & \mathbf{x}_{idT}^{(shock)T} 
\end{bmatrix} \right)
+  {\xi}^{(shock)*}_{idT},
\end{equation}
where the vector of independent variables can include both $\mathbf{x}_{idT}$, which represents the variables considered in the sample model, and $\mathbf{x}_{idT}^{(shock)}$, which covers independent variables not included in the sample data. Moreover, the influence of these variables on the logarithms of prices is described by the assumed function $h^{(shock)}$. Moreover, the random component ${\xi}^{(shock)*}_{idT}$ can be generated from different distribution than the random component in the sample model. 

Generating realizations of logarithms of prices based on (\ref{bootmodel_shock}) means that it is assumed that unanticipated shocks will occur in the future period with probability 1. This the reason the additional mixed bootstrap model is introduced. In this case future realizations of $\textbf{Y}_T$ are generated to obtain $\textbf{Y}^{(mix)*}_T$ with the $i$th element given by:
\begin{equation} \label{bootmodel_mix}
Y_{idT}^{(mix)} = p_{shock} \times Y_{idT}^{(shock)*} + (1- p_{shock}) \times Y_{idT}^{*}, 
\end{equation}
where $p_{shock}$ is the assumed probability of a shock, $Y_{idT}^{*}$ are generated based on (\ref{bootmodel_1T}) and $Y_{idT}^{(shock)*}$ are generated based on (\ref{bootmodel_shock}).

In this case, estimators of the $RMSE$ and $QAPE$ under model (\ref{bootmodel_shock}) (or model (\ref{bootmodel_mix})) are given by equations (\ref{RMSE}) and (\ref{QAPE}), where $\theta=\theta(G^{-1}(\textbf{Y}_T))$ 
and $\hat{\theta}=\hat{\theta}(G^{-1}(\textbf{Y}_S))=\theta(G^{-1}(\hat{\mathbf{Y}}_T))$ are replaced with:
\begin{itemize}
    \item $\theta^{*} = \theta(G^{-1}(\textbf{Y}^{(shock)*}_T))$ (or $\theta^{*} = \theta(G^{-1}(\textbf{Y}^{(mix)*}_T))$ for model (\ref{bootmodel_mix})) 
\item $\hat{\theta}^{*} = \hat{\theta}(G^{-1}(\textbf{Y}^{*}_S))=\theta(G^{-1}(\hat{\mathbf{Y}}^{*}_T))$.
\end{itemize}
Therefore, the bootstrap errors $\hat{\theta}^{*}  - \theta^{*}$ under models (\ref{bootmodel_shock}) and (\ref{bootmodel_mix}) account for unexpected shocks, because they are incorporated in the formula of $\theta^{*}$ (under  misspecified model addressing unanticipated shocks), despite they simultaneously remain unaccounted for in the formula of $\hat{\theta}^{*}$ which is not feasible in practice. In summary, the proposed approach allows for the estimation of the ex ante prediction accuracy under scenarios involving unanticipated shocks of forecasts, in which unexpected shocks cannot be accounted for.

It is worth noting that although in our proposal, the mixed model is used to generate logarithms of prices both for sample and future observations under scenarios without unanticipated shocks, it can be replaced by any model. Similarly, any modification of a general model (\ref{bootmodel_shock}) allowing for the unanticipated shocks can also be applied. In our opinion, not the specific formulae of these bootstrap models are crucial, but the idea of the generation sample values based on models where unanticipated shocks are not taken into account, and future values under a model where these shocks are included. 

\subsection{Results under unanticipated shocks versus reference values} \label{subsec:procedure_step3}

In subsection \ref{subsec:extended_approach}, we introduced a methodology for estimating ex ante prediction accuracy under market shocks. This methodology addresses two scenarios: the first, where a shock occurs with a specified probability, and the second, where it occurs with a probability of 1. These outcomes are of primary interest. In subsection \ref{subsec:traditional_approach}, we propose calculating the same measures in the absence of unanticipated shocks. These results will serve as reference values. By comparing these results, we can conduct a thorough analysis of the chosen prediction method's properties, not only under a model based on historical data but also in the context of unexpected market changes.

%\section{Real data application}
\section{USA market - the improved assessment of prediction} \label{sec:application}

We calculate forecasts of eight dispersion measures for the population and nine subpopulations under model (\ref{themodel}), which do not account for unanticipated shocks, and estimate ex ante prediction accuracy across nine scenarios. These include one scenario without unanticipated shocks (under model (\ref{themodel}) and eight shock scenarios. Based on each scenario, we estimate the ex ante prediction accuracy measures $RMSE$ and $QAPE$, which will allow us to observe changes in prediction accuracy due to the unanticipated shocks in the market. 

\subsection{Simulation scenarios}\label{subsec:Simulation scenarios}

The considered simulation scenarios used in the bootstrap procedures are as follows:
\begin{itemize}
    \item $s0$ -- scenario without unanticipated shocks (based on (\ref{bootmodel_1S}) and (\ref{bootmodel_1T})),
    \item $s1$ -- scenario with unanticipated shock due to reduction of government subsidies influencing green building taking place with the assumed probability denoted by $p_{shock}$, 
    \item $s11$ -- scenario  with shock scenario $s1$ assumed to take place in the future period with probability 1,
    \item $s2$ -- scenario with unanticipated shock due to introduced regulations against short-term rentals taking place with the assumed probability denoted by $p_{shock}$, 
    \item $s21$ -- scenario  with shock scenario $s2$ assumed to take place in the future period with probability 1,
     \item $s3$  -- scenario  with unanticipated shock due to the bursting of the real estate bubble  taking place with the assumed probability denoted by $p_{shock}$, 
    \item $s31$ -- scenario with shock scenario $s3$ assumed to take place in the future period with probability 1,
     \item $s4$ -- scenario with unanticipated shock due to a hurricane in the South and Northeast regions  taking place with the assumed probability denoted by $p_{shock}$, 
    \item $s41$ -- scenario with shock scenario $s4$ assumed to take place in the future period with probability 1 in both considered regions.
\end{itemize}
Details of scenarios $s1$, $s2$, $s3$, and $4$ are presented in Appendix \ref{App.C}. Scenarios $s11$, $s21$, $s31$, and $41$ are special cases of  $s1$, $s2$, $s3$, and $4$, where $p_{shock} = 1$. The assumed realisation of each scenario is inspired by historical external data, see Appendix \ref{App.C} for more details.

\subsection{Results}\label{subsec:Results}
%Let us consider the ex ante prediction accuracy of forecasts of the considered dispersion measures in the population and across the subpopulations. 

Determining the individual impact of a specific shock on the prediction accuracy of a particular dispersion measure is quite challenging.  This complexity arises from the combined effects of the shock's probability, the proportion of properties exposed to the shock, and the random fluctuations in prices. While one might expect that a shock would lead to an increase in the estimates of prediction accuracy measures, there are instances -- particularly when the price dispersion in certain subpopulations diminishes due to the shock -- where a decrease in these estimates can also occur. Furthermore, predicting relative dispersion measures, such as (\ref{CV}) - (\ref{DDR}), requires cautious treatment and detailed analysis, because even small changes in the denominator (e.g., due to the appearance of a shock) can lead to large changes in the value of the measure. Given the intricate nature of these price changes, a simulation procedure, presented in Section \ref{sec:improved}, is necessary. 

\begin{figure}[H]
    \centering
    \includegraphics[width=1\linewidth]{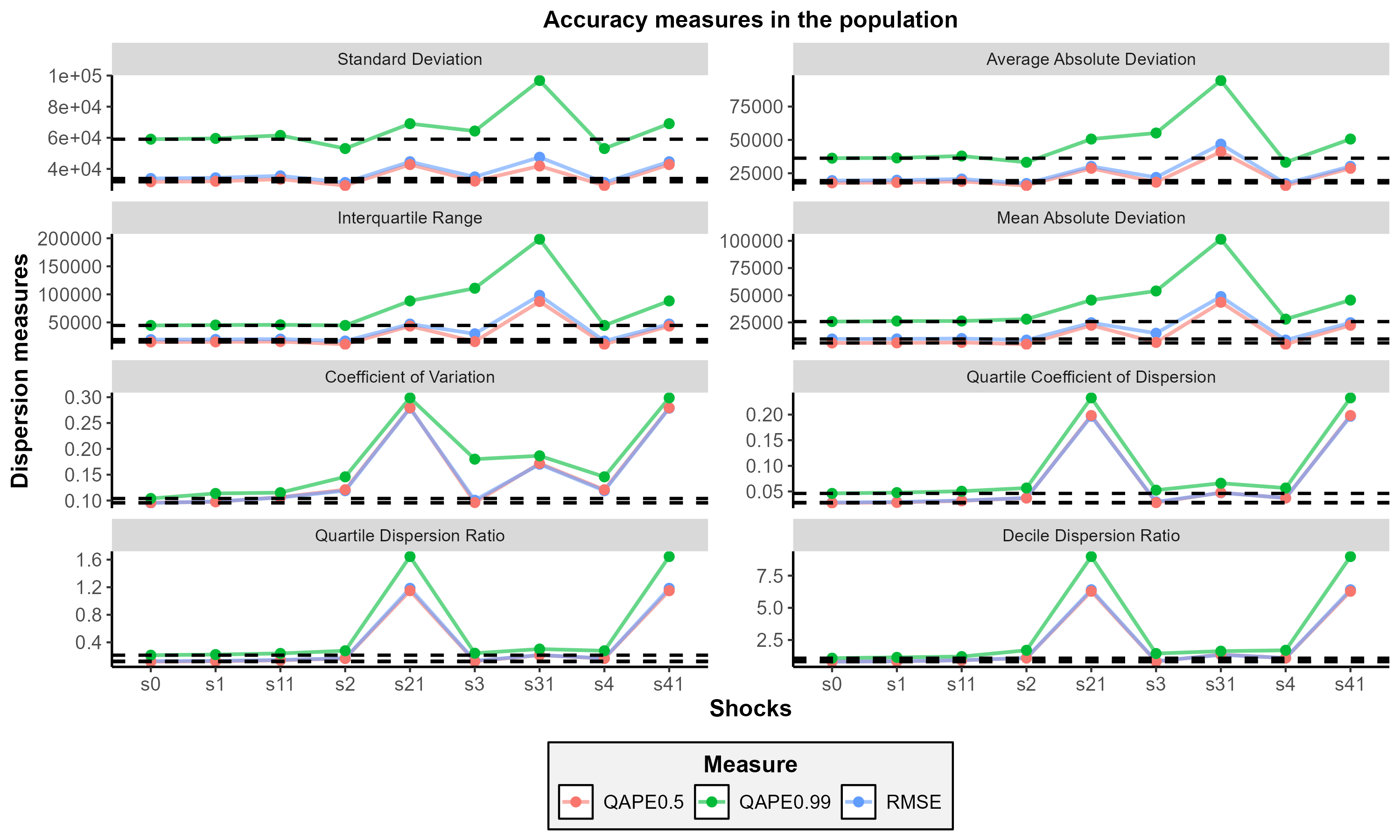}
    \caption{The comparison of prediction assessment of dispersion measures in the population according to shock presence. The dashed line represents the measures values for $s_0$ scenario.}
    \label{fig:mPop}
\end{figure}

In Figure \ref{fig:mPop}, we present a portion of the results obtained, focusing specifically on the prediction of dispersion measures for the entire population. The lines represent estimates of the $RMSE$ and $QAPE$  for orders 0.5 and 0.99, computed under scenarios  $s0$, $s1$, $s11$, $s2$, $s21$, $s3$, $s31$, $s4$, and $s41$ as displayed on the X-axis. Each subplot compares these ex ante accuracy measures in light of the various market shocks. The dashed black lines indicate the reference values obtained for the $s0$ scenario. Across each subplot, we observe similar patterns in the changes of the RMSE and $QAPE$s for orders $0.5$ and $0.99$. However, the values of the QAPE at order $0.99$ are notably higher, as this measure emphasizes the estimation of ex ante prediction accuracy within the upper tails of the distribution of absolute prediction errors. For scenarios $s1$ and $s11$, the values of these estimates of accuracy measures remain very close to the reference values, indicating the robustness of the prediction method under these market conditions. Conversely, for scenarios $s2$, $s3$, and $s4$, this performance is maintained only for certain dispersion measures, such as the Quartile Coefficient of Dispersion, Quartile Dispersion Ratio, and Decile Dispersion Ratio. In the cases of $s21$, $s31$, and $s41$, however, there are substantial departures from the respective reference lines.

\begin{figure}[H]
    \centering
    \includegraphics[width=1\linewidth]{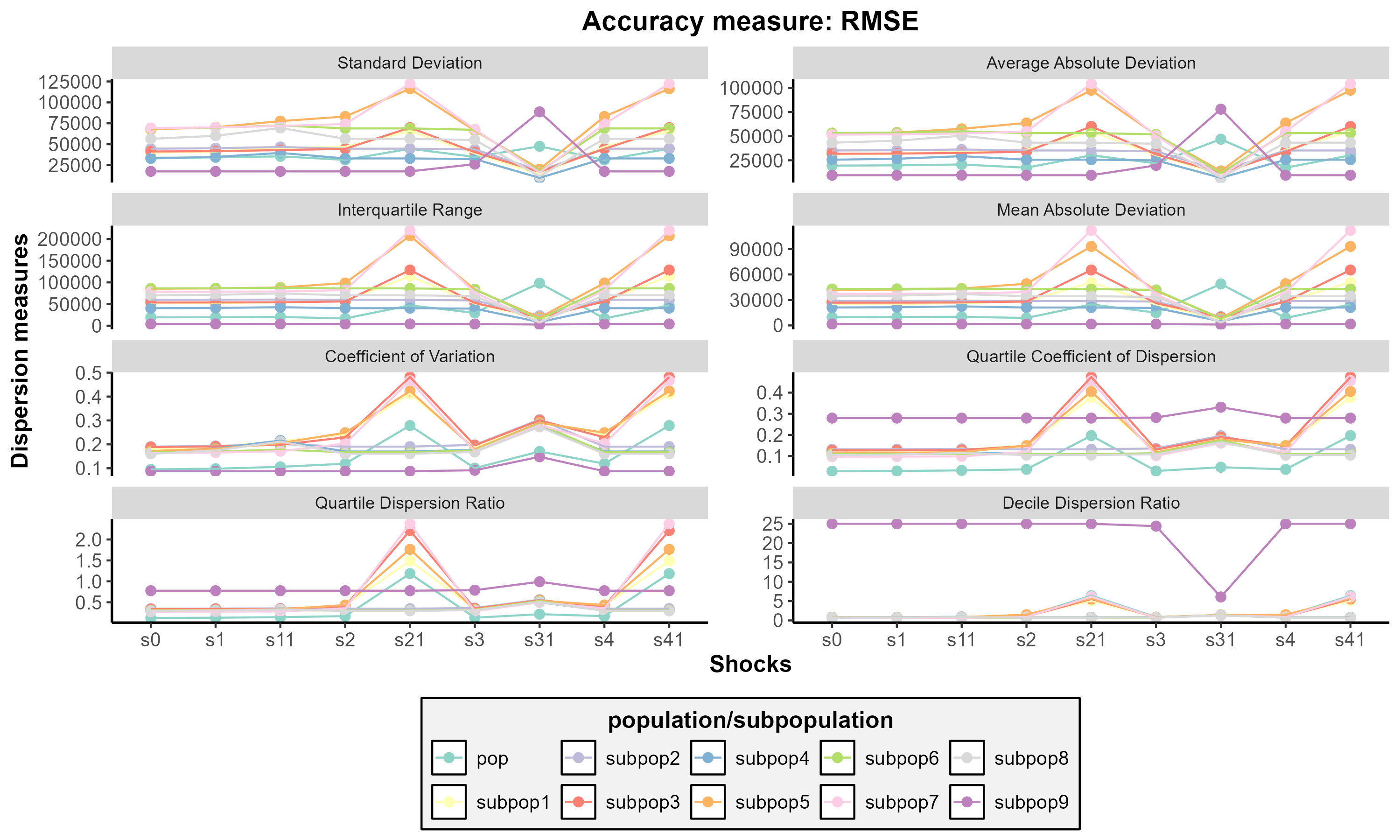}
    \caption{Estimates of the RMSE ex ante accuracy measure under variuos shock scenario}
    \label{fig:rmse}
\end{figure}

\begin{figure}[H]
    \centering
    \includegraphics[width=1\linewidth]{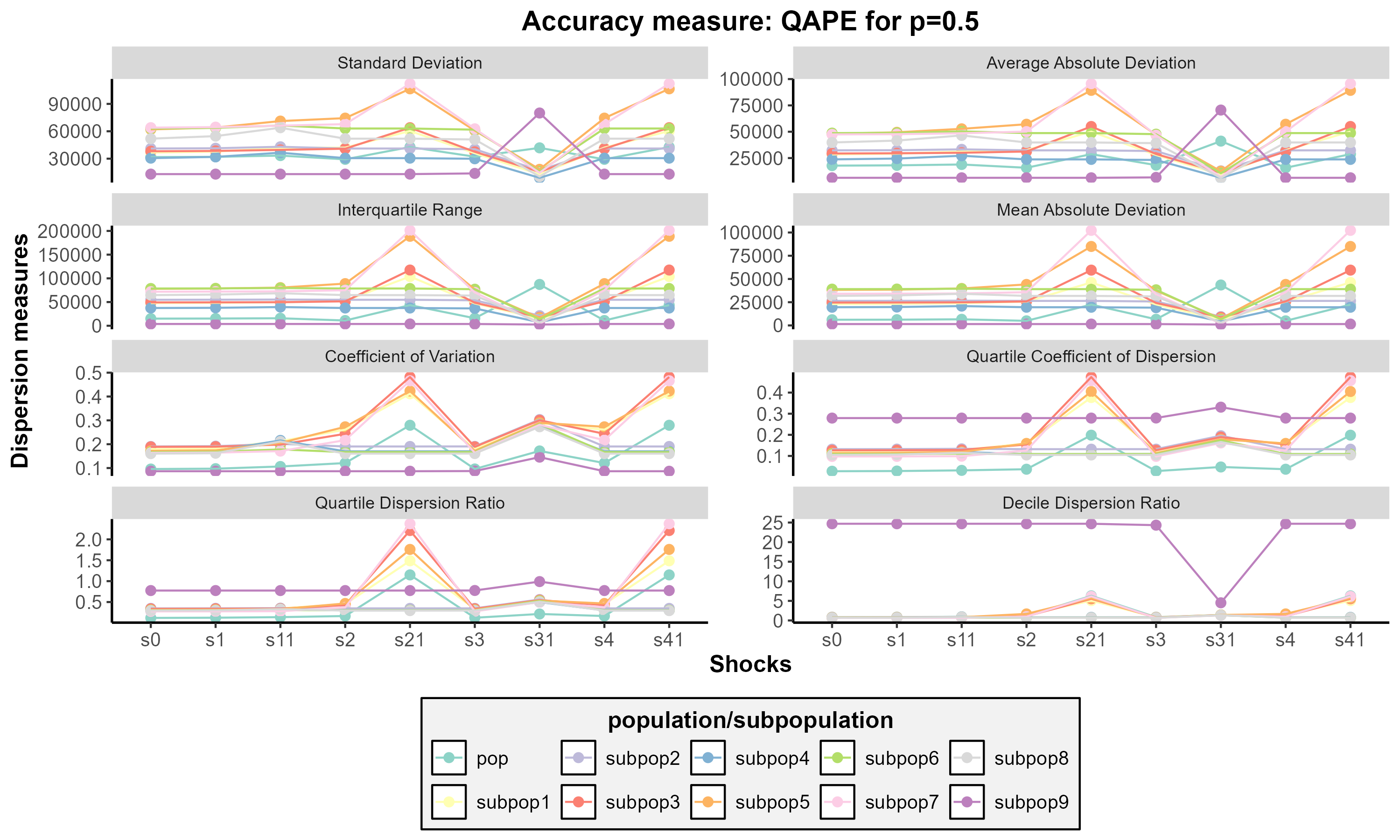}
    \caption{Estimates of the QAPE of order 0.5 ex ante accuracy measure under variuos shock scenario}
    \label{fig:qape5}
\end{figure}

\begin{figure}[H]
    \centering
    \includegraphics[width=1\linewidth]{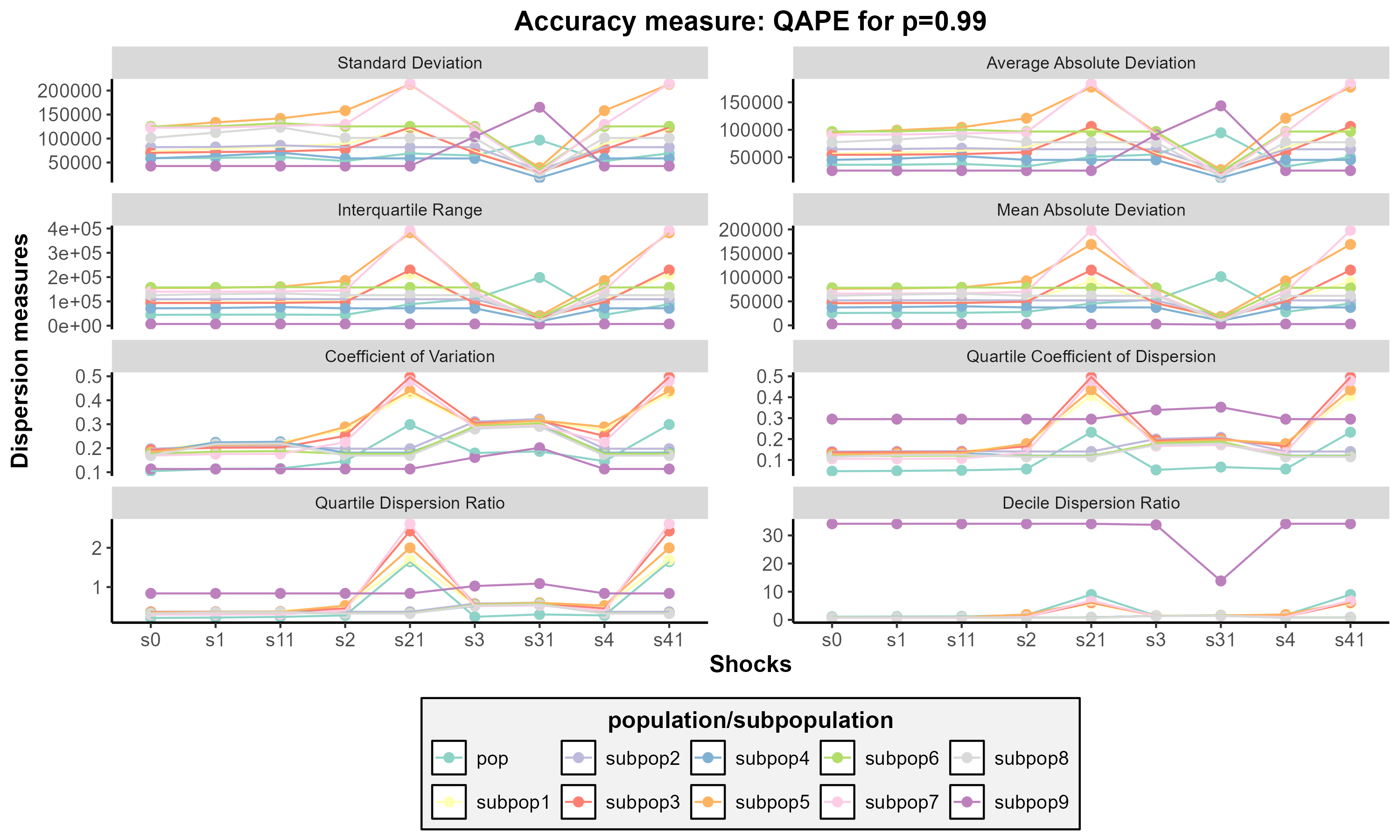}
    \caption{Estimates of the QAPE of order 0.99 ex ante accuracy measure under variuos shock scenario}
    \label{fig:qape99}
\end{figure}

%Let us consider the ex ante prediction accuracy of forecasts of the considered dispersion measures in the population and across the subpopulations. Figures \ref{fig:rmse}, \ref{fig:qape5}, and \ref{fig:qape99} illustrate the impact of various shocks ($s1$, $s11$, $s2$, $s21$, $s3$, $s31$, $s4$, $s41$) on the estimates of the RMSE, and QAPEs of orders 0.5 and 0.95 of these predictions in comparison with scenario without shocks on the market ($s0$). It is crucial to note, that in the considered shck scenarios the impact of a specific shock on the prediction accuracy of a forecast of a specific dispersion measure is very difficult to determine without the proposed procedure. It is because it is influenced by the joined effect of the probability of the shock, the percentage of the properties exposed to shock and the random change of the prices. Although it can be expected that the shock should imply the decrease in the estimated prediction accuracy, in some situations (where the dispersion of prices in specific subpopulation becomes smaller due to the shock), the increase of estimated accuracy can be observed as well. Due to this complex nature of changes, we need a simulation procedure proposed in Section \ref{sec:improved}.

Figures \ref{fig:rmse}, \ref{fig:qape5}, and \ref{fig:qape99} display all the results obtained, showing the estimates of $RMSE$ and $QAPE$ for orders 0.5 and 0.99 separately. This analysis not only focuses on predicting measures of population dispersion, as in the case of Figure \ref{fig:mPop}, but also examines subpopulations. Nevertheless, the general conclusions drawn are similar to those based on Figure \ref{fig:mPop}. It can be stated that shocks $s1$ and $s11$ (reduction of government subsidies with the assumed probability or with probability $1$) can be called neutral for the estimated ex ante prediction accuracy measures. Results for this shock scenarios show relative stability and minimal perturbation effects for all measures and all subpopulations. Shocks $s2$, $s3$ and $s4$ (regulations against short-term rentals, bursting of the real estate bubble, and a hurricane, all with given probability) have a moderate impact on the ex ante prediction accuracy. They usually lead to a substantial increase of estimated ex-ante accuracy measures compared with the $s0$ scenario. The same shock scenarios but occurring with probability 1 in the future period ($s21$, $s31$ and $s41$) have a high impact on the estimates of the considered ex ante prediction accuracy measures leading to large spikes visible in Figures \ref{fig:rmse}, \ref{fig:qape5}, and \ref{fig:qape99}. For most of the considered cases subpopulations respond differently to shocks scenarios. For example, subpopulation 9 often exhibits heightened sensitivity to particular shocks, while subpopulation 6 and 8 usually show relatively stable and low values across all measures and sceanarios. An interesting example is shock $s21$, which causes pronounced instability, affecting specific subpopulations more intensely. 

Based on the results, recommendations on the usage of the chosen forecasting method can be proposed. In the considered case, it can be argued that the proposed forecasting method is applicable not only under the assumed model (for scenario $s0$) but also under unanticipated shock scenarios such as $s1$, $s11$, $s2$, $s3$, and $s4$. However, if shock scenarios like $s21$, $s31$, or $s41$ occur in the future, we may observe an unacceptable impact on prediction accuracy.

% \begin{figure}
%     \centering
%     \includegraphics[width=1\linewidth]{QAPE5plot.png}
%     \caption{QAPE 0.5 accuracy measure}
%     \label{qape5}
% \end{figure}

% \begin{figure}
%     \centering
%     \includegraphics[width=1\linewidth]{QAPE95plot.png}
%     \caption{QAPE 0.95 accuracy measire}
%     \label{qape95}
% \end{figure}

% \begin{figure}
%     \centering
%     \includegraphics[width=1\linewidth]{QAPE99plot.png}
%     \caption{QAPE 0.99 accuracy measure}
%     \label{qape99}
% \end{figure}

\section{Conclusion}\label{subsec:Conclusion}
In the paper, the problem of prediction of dispersion measures of real estate prices on the U.S. market is considered. It highlights the complexity of forecasting in the presence of shocks, especially unanticipated shocks with unknown impact on the prices, posing challenges for accurate predictions. It proposed an improved, more nuanced approach to prediction accuracy estimation. Traditionally, when researchers estimate ex ante accuracy without the proposed improvement, they base their assessment on historical data, assuming that future performance of the variable will follow past changes. 
 The proposed method, however, offers additional insights across various shock scenarios. This allows for a more comprehensive analysis of how different potential market shocks affect the ex ante prediction accuracy for forecasting various dispersion measures, both for the entire population and its subpopulations. In the considered application of the proposed method, it was shown on which unanticipated shocks on the market the forecast is robust, and on which it is not. 

Our study has also some limitations related to the assumptions made in the prediction model and the handling of unanticipated shock scenarios.  However, the model used in the proposed method can be substituted with any other model. Additionally, the assumptions regarding shock scenarios can be modified or expanded to include any scenario, whether it is based on external data or shaped by the researcher's subjective beliefs.

Although these results and the proposed method can be particularly useful for analysts of the real estate market, they can be used in the prediction process of any phenomenon possibly affected by unexpected changes. It provides an opportunity to improve the prediction process, leading to a deeper understanding of the influence of shock scenarios on the future realizations of the variable of interest. Therefore, decision-makers can benefit from these insights by adjusting their decisions due to a more informed decision-making process. 

\section*{Acknowledgement}
Co-financed by the Minister of Science under the "Regional Initiative of Excellence" programme.

\begin{appendices}

\section{Model selection}\label{App.A}

The selection of the optimal combination of independent variables is achieved through the following procedure.
First, each potential independent variable is evaluated in both its original and log-transformed forms. The form selected for subsequent analysis is based on the absolute value of its correlation coefficient with LOG.PRICE. 
Then, variables with a Pearson correlation coefficient absolute value with LOG.PRICE of less than 0.4 are removed, except for those intended to be binary (0-1 variables). After this step, 47 potential auxiliary variables remain. 
After that, The correlation between potential auxiliary variables is analyzed. If the absolute value of the Pearson correlation between two variables exceeds 0.8, one of the variables is removed from the dataset. After this step, 15 potential continues auxiliary variables remain. Up to this point, qualitative variables, which will be employed as 0-1 variables later, are not taken into account.
In the next step, linear models are constructed using all possible combinations of the potential auxiliary variables, at this step including qualitative variables taken into account as sets of 0-1 variables. The model with the highest value of the adjusted R2 is selected.
Finally, the significance of each variable is tested using the permutation version of the Student's t-test assuminga significance level of $0.05$. The number of iterations equals $200$. This includes testing each level of qualitative variables, which have been transformed into binary (0-1) quantitative variables. In each step, we exclude the variable with the highest p-value, provided it is greater than 0.05. This procedure continues until all variables included in the model are significant at the 0.05 level.

As the result, 21 independent variables, including binary variables for different levels of variables of qualitative variables, are selected. The independent variables are:
\begin{itemize}
    \item LOG.SQFT - logarithms of the square footage of the home,
    \item HOUSEHOLDS - the number of households with one or more people 65 years and over in the region in the previous year (in millions),
    \item LOG.OWNER - logged number of the owner-occupied housing units with a mortgage with no real estate taxes paid in the region in the previous year,
    \item INCOME - median income in the past 12 Months of moved from abroad in USD in the region in the previous year (in 2022 inflation-adjusted dollars),
     \item LOCATION - where a home is placed: inside manufactured home communities, outside manufactured home communities, nonapplicable cases or for disclosure purposes,
    \item TITLED - how a home is titled: real estate, personal property, not titled, nonapplicable cases or for disclosure purposes,
    \item BEDROOMS - number of bedrooms: collapsed to two or fewer and three or more,
    \item SECURED - how a home is secured, if not on permanent masonry/concrete: tie down straps or other, not secured, nonapplicable cases or for disclosure purposes
    \item zero-one variables for each year,
    \item PPI - US producer price index,
    \item FED - monthly federal funds effective rate in percents.
\end{itemize}
LOCATION, TITLED, SECTIONS, BEDROOMS, SECURED, and REGION are qualitative variables and they are transformed to zero-one variables for modelling purposes. 

\section{House prices and house areas}\label{App.B}

\begin{figure}[ht!] 
    \centering
\includegraphics[scale = 0.78]{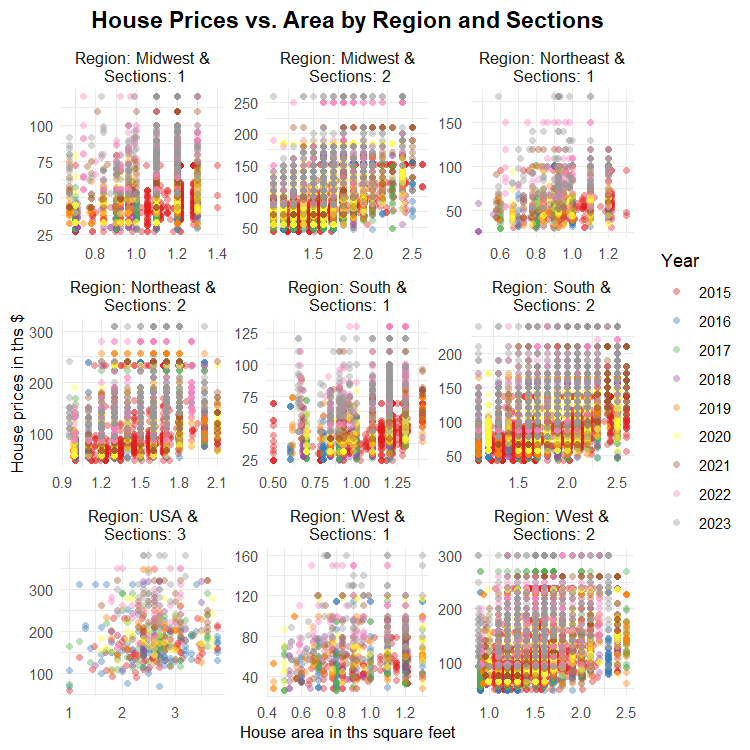}
\caption{Scatterplots of house prices and house areas in subpopulations in years 2015--2023}
\label{scatterplots}
\end{figure}

\newpage
\section{Shock scenarios}\label{App.C}

The first shock scenario ($s1$) is due to the assumed reduction of government subsidies influencing green building in the future period. It is inspired by the data from the U.S. Energy Information Administration (i.e. Residential Energy Consumption Survey and Commercial Energy Consumption Survey). The probability of the shock is assumed to be $p_{shock} = 0.25$, and it is assumed to cause a random decrease in the price of only some properties (influenced by green building regulations): 2.5\% of properties in the Northeast region, 0.8\% of properties in the Midwest region, 0.6\% in the South region, and 3.8\% in the West region. The average decrease is assumed to be 17.5\% with a standard deviation equal to one third of the mean, and it is assumed to follow a normal distribution.

The second shock scenario ($s2$) is referred to as regulations against short-term rentals. It is adopted based on the data from Lisbon, one of the pioneering cities in implementing regulations against short-term rentals, where restrictions in some areas were introduced in 2018 and their influence on real estate prices has already been assessed. It is assumed that with a probability of $p_{shock} = 0.9$, these regulations will affect real estate prices in the future period, but only for 4.6\% of properties. For properties with two or fewer bedrooms, prices are expected to be smaller by an average of 11.934\% compared to prices generated by models that do not account for this shock. This decrease follows a normal distribution, with a standard deviation set to one-third of the mean. Similarly, for properties with three or more bedrooms, prices are projected to decline by an average of 2.4\%, also following a normal distribution with a standard deviation of one-third of the mean.

The third shock scenario, denoted by $s3$, is referred to as the bursting of the real estate bubble. This scenario is based on the historical data from the U.S. Census Bureau and U.S. Department of Housing and Urban Development data on the annual median sales price of houses sold for the United States. We assume that this shock occurs in future period with a probability of $p_{shock} = 0.05$, resulting in an average decrease of all property prices by $6.88\%$ compared to prices generated by models that do not account for this shock. To model this shock, we use a normal distribution with a standard deviation of $2.293\%$, which is one-third of the mean. This approach ensures that prices for nearly all properties will randomly decrease, although it is acceptable for some properties to experience slight increases in rare cases.

The fourth exogenous shock on real estate market, referred to as $s4$, involves a hurricane occurring in the future period in the South and Northeast regions. This scenario is inspired by historical data collected by the Hurricane Research Division of the Atlantic Oceanographic and Meteorological Laboratory since the beginning of the 20th century (from 124 years). It includes three sub-scenarios that simultaneously affect property prices:
\begin{itemize}
    \item With a probability of $p_{shock} = 0.02$, the prices of 30\% of properties in the Northeast region are decreased, compared to prices generated by models that do not account for the shock, by random values ranging from 6\% to 16\% (using a uniform distribution).
    \item With a probability of \(p_{shock} = 0.69\), the prices of 38\% of properties in the South region are decreased—again compared to prices generated by models not covering the shock—by random values from 0.5\% to 3.8\% (also using a uniform distribution).
    \item With a probability of \(p_{shock} = 0.08\), the prices of 30\% of properties in the Northeast region are decreased by random values from 6\% to 16\%, while simultaneously, the prices of 38\% of properties in the South region are decreased by random values from 0.5\% to 3.8\% (both following a uniform distribution).
\end{itemize}
Hence, in the simulation setup for this case, which consists of $B = 2000$ iterations, the distribution of iterations is as follows: $0.02B$ iterations are conducted according to the first sub-scenario, $0.69B$ iterations according to the second sub-scenario, $0.08B$ iterations according to the third sub-scenario, and the remaining $0.21B$ iterations follow the scenario without shocks.

% An appendix contains supplementary information that is not an essential part of the text itself but which may be helpful in providing a more comprehensive understanding of the research problem or it is information that is too cumbersome to be included in the body of the paper.

% %%=============================================%%
% %% For submissions to Nature Portfolio Journals %%
% %% please use the heading ``Extended Data''.   %%
% %%=============================================%%

% %%=============================================================%%
% %% Sample for another appendix section			       %%
% %%=============================================================%%

% %% \section{Example of another appendix section}\label{secA2}%
% %% Appendices may be used for helpful, supporting or essential material that would otherwise 
% %% clutter, break up or be distracting to the text. Appendices can consist of sections, figures, 
% %% tables and equations etc.

\end{appendices}

% %%===========================================================================================%%
% %% If you are submitting to one of the Nature Portfolio journals, using the eJP submission   %%
% %% system, please include the references within the manuscript file itself. You may do this  %%
% %% by copying the reference list from your .bbl file, paste it into the main manuscript .tex %%
% %% file, and delete the associated \verb+\bibliography+ commands.                            %%
% %%===========================================================================================%%

\bibliography{snarticle}% common bib file

\end{document}